\documentclass[letterpaper,10 pt,conference]{ieeeconf}  
\IEEEoverridecommandlockouts                 
\usepackage{cite}
\usepackage{amsmath,amssymb,amsfonts}
\usepackage{graphicx}
\usepackage{hyperref}
\usepackage{textcomp}
\usepackage{mathbbol}
\usepackage{dirtytalk}              
\usepackage{graphicx}
\usepackage{amsmath}
\usepackage{mathtools}
\usepackage{verbatim}
\usepackage{amssymb}
\usepackage{epstopdf}

\newtheorem{definition}{Definition}
\newtheorem{remark}{Remark}
\newtheorem{thm}{Theorem}
\newtheorem{lemma}{Lemma}

\newtheorem{expm}{Example}

\usepackage[font=footnotesize,labelfont=bf]{caption}
\usepackage{subcaption}

\title{\LARGE \bf
Opinion Clustering under the Friedkin-Johnsen Model: Agreement in Disagreement}
\author{Aashi Shrinate$^{1}$, \IEEEmembership{Student member, IEEE} and Twinkle Tripathy$^2$, \IEEEmembership{Senior Member, IEEE}
\thanks{$^{1}$Aashi Shrinate is a research scholar and $^{2}$ Twinkle Tripathy is an Assistant Professor in the Control and Automation specialization of the Department of Electrical Engineering, Indian Institute of Technology Kanpur, Kanpur, Uttar Pradesh, India, 208016. Email: {\tt\small aashis21@iitk.ac.in and ttripathy@iitk.ac.in}.}
}
\begin{document}
\maketitle
\thispagestyle{empty}
\pagestyle{empty}
\begin{abstract}
The convergence of opinions in the Friedkin-Johnsen (FJ) framework is well studied, but the topological conditions leading to opinion clustering remain less explored. To bridge this gap, we examine the role of topology in the emergence of opinion clusters within the network. The key contribution of the paper lies in the introduction of the notion of topologically prominent agents, referred to as Locally Topologically Persuasive (LTP) agents. Interestingly, each LTP agent is associated with a unique set of (non-influential) agents in its vicinity. Using them, we present conditions to obtain opinion clusters in the FJ framework in any arbitrarily connected digraph. A key advantage of the proposed result is that the resulting opinion clusters are independent of the edge weights and the stubbornness of the agents. Finally, we demonstrate using simulation results that, by suitably placing LTP agents, one can design networks that achieve any desired opinion clustering. 
\end{abstract}

\section{INTRODUCTION}
In recent decades, social media networks have become a major platform for advertising, conducting socio-political campaigns, and even spreading misinformation. This evolution has resulted in an interest towards understanding the impact of social interactions on the formation of opinion. The popular opinion dynamics, such as DeGroot model \cite{degroot1974consensus}, the FJ model \cite{friedkin1990opinions}, the Hegselmann-Krause model \cite{rainer2002opinion}, and the Biased assimilation model \cite{dandekar2013biased}, capture the complex aspects of human interactions. Among these, the FJ model is popular because it captures a diverse range of emergent behaviours. Its popularity also stems from its analytical tractability and performance over large datasets \cite{Community_Cleavage}. 

The FJ model was proposed in \cite{friedkin1990opinions} to justify the \textit{disagreement} among closely interacting individuals. The opinions evolving under the FJ model converge to disagreement due to the presence of biased individuals, who are commonly known as the stubborn agents. In \cite{friedkin1990opinions}, the authors show that if each non-stubborn agent in the network has a directed path from a stubborn agent, the opinions converge asymptotically at a steady state. Further, they demonstrate that the final opinions depend on the initial opinions of only the stubborn agents and lie in the convex hull of these opinions. 
 The authors in \cite{parsegov2016novel} introduced the notion of \textit{oblivious agents}, who are non-stubborn and do not have a path from any stubborn agent. The algebraic and graph-theoretic conditions that ensure convergence in networks having oblivious agents were presented in \cite{parsegov2016novel} and \cite{TIAN2018213}, respectively. Additionally, in \cite{TIAN2018213}, the authors show that the final opinions depend on the initial opinions of both the stubborn agents and some oblivious agents. 

In opinion dynamics, emergent behaviours such as \textit{consensus} and \textit{opinion clustering} are of keen interest to explain and achieve a desired social outcome. While consensus is being widely explored even in the FJ framework \cite{LWang_CFJ}, this work focuses on {opinion clustering}. A network achieves opinion clustering when there is a group of two or more agents such that, for any initial opinions, the final opinions of agents in the group are equal. Opinion clustering represents swarm behaviour \cite{richardson2022two} and has applications in task distribution \cite{bizyaeva2022nonlinear} and formation control \cite{anderson2008rigid}. The works \cite{yu2010group,xia2011clustering,multiconsensus} present graph-theoretic conditions under which opinion clustering occurs in continuous-time linear invariant systems. 
While it is known that disagreement occurs in the FJ framework,
the conditions for opinion clustering remain less explored. In \cite{yao2022cluster}, the authors present the graph-theoretic conditions to achieve opinion clustering in the FJ framework in networks with oblivious agents. However, the results apply only to networks where each cycle contains only one stubborn agent.

Contrary to \cite{yao2022cluster}, our work aims to explain the emergence of opinion clusters in the FJ framework in \textit{arbitrarily connected digraphs}. In opinion dynamics, an opinion cluster at steady state represents a group of individuals with a shared view on a topic. Our focus is specifically on how the network topology directs a certain group of agents 
to form an opinion cluster. With this, we now highlight our major contributions as follows:
\begin{itemize}
    \item We define a new type of agent called the \textbf{\textit{ Locally Topologically Persuasive}} (LTP) agent. We show that each LTP agent is associated with a unique set of non-influential agents in its vicinity.
    
    \item We establish that an LTP agent and the agents associated with it always form an opinion cluster under the FJ model. This result generalises the topology-based conditions presented in \cite{yao2022cluster} to arbitrarily connected digraphs.
\item Unlike \cite{yao2022cluster}, the proposed LTP-based conditions for opinion clustering show that even a group of non-influential agents can form an opinion cluster
    
    \end{itemize}
The organisation of the paper is as follows: Sec. \ref{Sec2} introduces the notations and the relevant preliminaries. Sec. \ref{Sec:FJ} motivates the analysis of the topological dependence of opinion clusters. Sec. \ref{sec:Kron_Red} presents a framework to relate final opinions of any set of agents in the graph. We introduce the LTP agents and present the conditions for opinion clustering in Sec. \ref{sec:OC}. Finally, we conclude in Sec. \ref{Sec:conc}. 

\section{Notations and Preliminaries}
\label{Sec2}
\subsection{Notations}
\label{subsec:NOT}
$\mathbb{1}$ ($\mathbb{0}$) denotes a vector/matrix
of appropriate dimensions with all entries equal to $+1\ (0)$. $I$ denotes the identity matrix of appropriate dimension. 
$M=$ diag$(m_1,m_2,...,m_n)$ is a diagonal matrix with entries $m_1,...,m_n$. A set $\{1,2,...,k\}$ is denoted by $[k]$. The spectral radius of a matrix $M$ is given by $\rho(M)$.
\subsection{Graph Preliminaries}
A graph is defined as $\mathcal{G}=(\mathcal{V},\mathcal{E})$ with $\mathcal{V}=\{1,2,...,$ $n\}$ denoting the set of $n$ agents and $\mathcal{E} \subseteq \mathcal{V} \times \mathcal{V}$ giving the set of directed edges in the network. An edge $(i,j) \in \mathcal{E}$ 
informs that node $i$ is an in-neighbour of node $j$ and node $j$ is an out-neighbour of node $i$.
%
A \textit{source} is a node without any in-neighbours.
The matrix $W=[w_{ij}] \in \mathbb{R}^{n \times n}$ is the weighted adjacency matrix of graph $\mathcal{G}$ with $w_{ij} > 0$ only if $(j,i) \in \mathcal{E}$, otherwise $w_{ij}= 0$. The in-degree of an agent $i\in \mathcal{V}$ is defined as $d_{in}(i)=\sum_{j=1}^{n}w_{ij}$. The Laplacian matrix is defined as $L=D-W$, where $D=$ diag$(d_{in}(1),...,d_{in}(n))$. The loopy Laplacian matrix is defined as $Q =L+\operatorname{diag}(w_{11},w_{22},...,w_{nn})$, where $w_{ii}$ is the weight of self loop at agent $i\in[n]$. If a network does not have any self-loops, then $Q=L$ and is called a loopless Laplacian matrix.

A \textit{walk} is an ordered sequence of nodes such that each pair of consecutive nodes forms an edge in the graph. A \textit{cycle} is a walk whose initial and final nodes coincide. If no nodes in a walk are repeated, it is a \textit{path}. A graph is \textit{aperiodic} if the lengths of all its cycles are coprime. An undirected graph is \textit{connected} if a path exists between every pair of nodes. A directed graph (digraph) is a \textit{strongly connected} graph if a directed path exists between every pair of nodes in the graph. A digraph is \textit{weakly connected} if it is not strongly connected, but its undirected version is connected. A maximal subgraph of $\mathcal{G}$ which is strongly connected forms a \textit{strongly connected component} (SCC). An SCC is an \textit{independent strongly connected component} (iSCC) if every node in the SCC has all its in-neighbours within the same SCC. 


\subsection{Kron reduction}
\label{subsec:KR}
Let $M\in \mathbb{R}^{p \times p}$ and $\alpha,\beta \subseteq [p]$ be index sets. Then $M[\alpha,\beta]$ is a submatrix of $M$ with rows indexed by $\alpha$ and columns indexed by $\beta$. The submatrix $M[\alpha]:=M[\alpha,\alpha]$ and $\alpha^{c}:=[p] \setminus \alpha$. If $M[\alpha^c]$ is non-singular, then the Schur complement of $M[\alpha^c]$ in $M$ is defined as:
\begin{align}
\label{eqn:Schur_complement}
    M/\alpha^c=M[\alpha]-M[\alpha,\alpha^c](M[\alpha^c])^{-1}M[\alpha^c,\alpha]
\end{align}

Under Kron reduction, the Schur complement of the loopy Laplacian matrix $Q$ is determined. 

\begin{lemma}[\hspace{-0.01cm}\cite{Kron_red_digraphs}]
\label{lm:Basic_properties}
Consider a loopy Laplacian matrix $Q \in \mathbb{R}^{p}$ and $\mathcal{G}(Q)$ be the digraph associated with $Q$. Let $\alpha \subset [p]$.
\begin{enumerate}
    \item $Q/\alpha^c$ is well defined if, for each node $i\in \alpha^c$, there is a node $j\in \alpha$ such that a path $j \to i$ exists in $\mathcal{G}(Q)$.  \footnote{\label{foot:1}(Note that this condition is adapted to our framework because we consider $w_{ij}>0$ when $(j,i)\in \mathcal{E}$.)}
  
    \item If $Q$ is a loopless Laplacian matrix, then $Q/\alpha^c$ is also a loopless Laplacian matrix.
    
\end{enumerate}
\end{lemma}

\section{FJ model and emergent behaviours}
\label{Sec:FJ}
Consider a network $\mathcal{G}$ of $n$ agents with vector $\mathbf{x}(k)=[x_1(k),...,x_n(k)]$ denoting the opinions of the agents at the $k^{th}$ instance.  Under the FJ model, the opinions of agents evolve as follows: 
\begin{align}
    \mathbf{x}(k+1)=(I-
    \beta)W \mathbf{x}(k)+\beta \mathbf{x}(0)
    \label{eq:FJ_opinion_dynamics}
\end{align}
where $\beta=$ diag$(\beta_1,...,\beta_n)$ is a diagonal matrix with each $\beta_{i}\in [0,1]$
representing the stubbornness of agents. An agent $i\in \mathcal{V}$ is stubborn if $\beta_i>0$. $W$ is row-stochastic.

Suppose $\mathcal{G}$ is weakly connected and there is a non-stubborn agent that does not have a directed path from any stubborn agent. In \cite{parsegov2016novel}, such agents are called \textit{oblivious agents}. Without loss of generality, we can renumber the nodes in $\mathcal{G}$ such that the nodes $\{1,2,...,n_o\}$ are the oblivious agents and the rest are the non-oblivious, where $n_o\in[n]$. Additionally, we consider the oblivious agents that form an iSCC to be grouped together. Now, we can re-write the FJ-model \eqref{eq:FJ_opinion_dynamics} as:
\begin{align*}
\mathbf{x}_1(k+1)&=W_{11}\mathbf{x}_1(k) \nonumber\\
\mathbf{x}_{2}(k+1)&=(I-\bar{\beta})\big(W_{21}\mathbf{x}_1(k)+ W_{22}\mathbf{x}_2(k)\big)+\bar{\beta}\mathbf{x}_{2}(0) 
\end{align*}

where $\mathbf{x}_1(k)\in \mathbb{R}^{n_o}$ and $\mathbf{x}_2(k)\in\mathbb{R}^{(n-n_o)}$ denote the opinions of oblivious and non-oblivious agents. The matrix $W$ gets partitioned as $W=\begin{bmatrix}
    W_{11} & \mathbb{0} \\ W_{21} & W_{22}
\end{bmatrix}$ and $\bar{\beta} \in \mathbb{R}^{(n-n_o)\times (n-n_o)}$ is a diagonal matrix with entries equal to the stubbornness of the non-oblivious agents. 

\begin{lemma}[\cite{parsegov2016novel,TIAN2018213}]
\label{lm:final_op}
Consider a weakly connected network $\mathcal{G}$ with opinions evolving under the FJ model \eqref{eq:FJ_opinion_dynamics}.
\begin{itemize}
    \item If $\mathcal{G}$ does not contain any oblivious agents, the final opinions $\mathbf{x}^*$ always converge to 
    \begin{align*}
   \mathbf{x}^{*}=(I-(I-
    \beta)W)^{-1} \beta \mathbf{x}(0)   
\end{align*}

\item If $\mathcal{G}$ has oblivious agents, then the opinions converge at steady state only if each iSCC composed of two or more oblivious agents is aperiodic. The final opinions $\mathbf{x}^*$ converge to
 \begin{align*}
  \mathbf{x}_1^*&= W_{11}^* \mathbf{x}_1(0) \nonumber \\
 \mathbf{x}_2^*&=(I-(I-\bar{\beta})W_{22})^{-1} \big( (I-\bar{\beta}) W_{21}W_{11}^* \mathbf{x}_1(0) \nonumber\\
 &+ \bar{\beta}\mathbf{x}_2(0)\big)
\end{align*}

where 
$W_{11}^*=\lim_{k \to \infty}W_{11}^k$.
\end{itemize}
\end{lemma}
    
Since $\mathbf{x}^*$  depends only on the initial opinions of the stubborn and the oblivious agents in iSCCs, these agents are called the \textit{influential agents}. $\mathcal{I}$ denotes the set of all influential agents in $\mathcal{G}$. From Lemma \ref{lm:final_op}, it follows that the impact of these influential agents on $\mathbf{x}^*$ depends on two factors: the network topology (captured by $W$) and the stubbornness of agents. 

In general, the final opinions converge to different values, causing disagreement among agents. A special case of disagreement occurs when a group of two or more agents converges to the same final opinion and forms an opinion cluster. Mathematically, an opinion cluster is defined as:
\begin{definition}  A set of agents $\mathcal{C}\subset \mathcal{V}$ form an \textbf{opinion cluster} if for any initial conditions $\mathbf{x}(0)\in \mathbb{R}^{n}$
\begin{align*}
    x_i^*-x_j^*=0 \quad \forall \ i,j\in \mathcal{C}
\end{align*}

where $x_k^*$ denotes the final opinion of agent $k\in \mathcal{V}$. 
\end{definition}

\begin{figure}[h]
    \centering
    \begin{subfigure}{0.21\textwidth}
        \centering
    \includegraphics[width=0.75\linewidth]{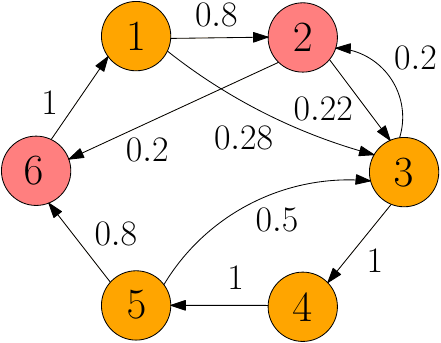}
    \caption{Network $\mathcal{G}$}
    \label{Fig:expm_1_graph}
    \end{subfigure}
    \hfill
        \begin{subfigure}{0.25\textwidth}
        \centering
    \includegraphics[width=0.9\linewidth]{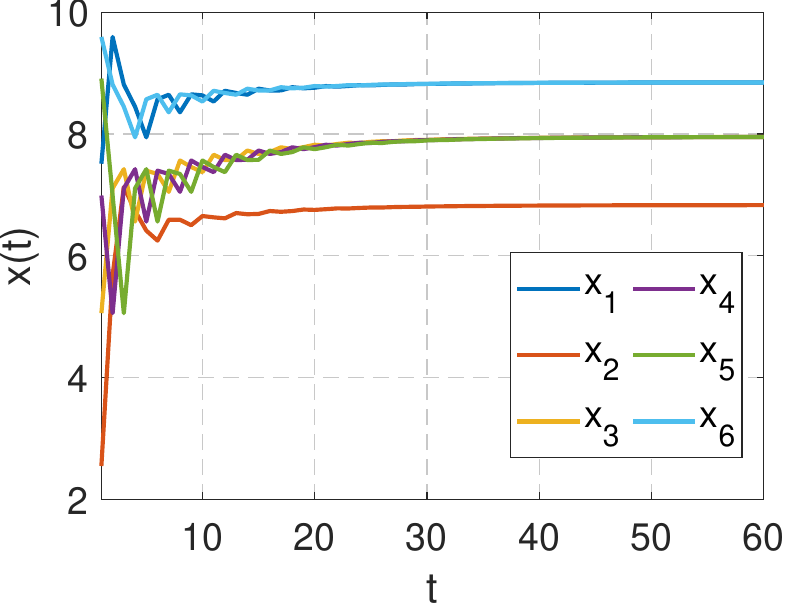}
    \caption{Evolution of opinions $\mathcal{G}$.}
       \label{Fig:expm_1_sim}
    \end{subfigure}
\begin{subfigure}{0.21\textwidth}
        \centering
    \includegraphics[width=0.75\linewidth]{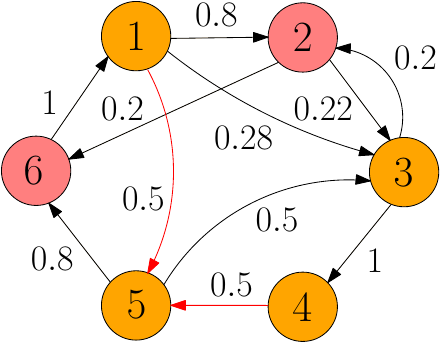}
    \caption{Modified network $\mathcal{\hat{G}}$}
    \label{Fig:modified_g}
    \end{subfigure}
    \hfill
        \begin{subfigure}{0.25\textwidth}
        \centering
    \includegraphics[width=0.9\linewidth]{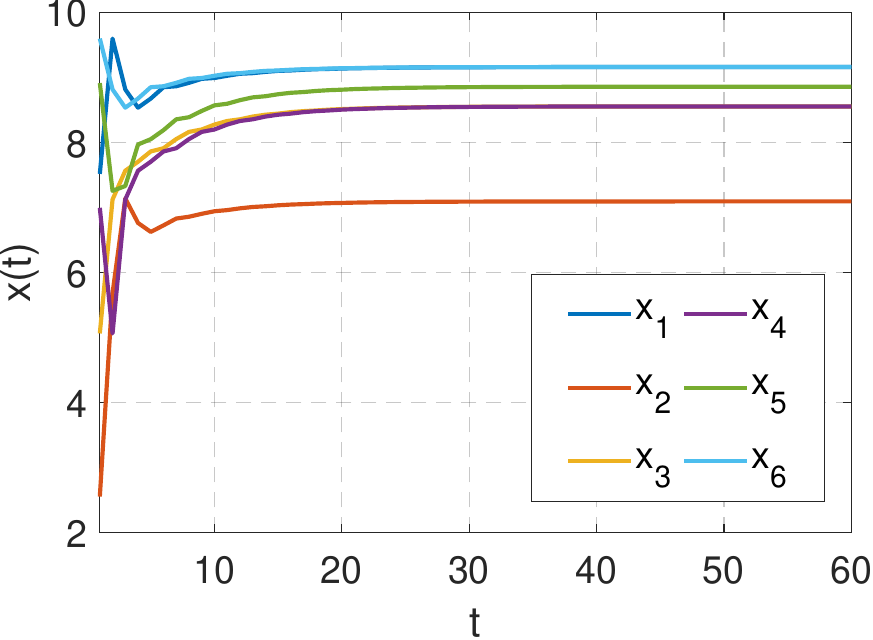}
    \caption{Evolution of opinions in  $\mathcal{\hat{G}}$}
       \label{Fig:final_op_modified_g}
    \end{subfigure}
      
    \caption{Simple modification to $\mathcal{G}$ affects the opinion clusters in final opinion.
    Throughout the paper, we represent the stubborn agents and the non-stubborn agents by red nodes and orange nodes, respectively. The modified edges in $\mathcal{\hat{G}}$ are highlighted in red.}
    \label{fig:placeholder}
\end{figure}

Naturally, the question arises as to how the formation of these opinion clusters depends on network topology and stubborn behaviour. The following example illustrates the effect of network topology on the opinion clusters. 
\begin{expm}
\label{Example-1}
  Consider the network $\mathcal{G}$ shown in Fig. \ref{Fig:expm_1_graph}. $\mathcal{G}$ has two stubborn agents: $2$ and $6$ with $\beta_2=0.3$, and $\beta_6=0.6$, and no oblivious agents. For initial opinions chosen from a uniform distribution on $[0,10]$, three opinion clusters form under the FJ model \eqref{eq:FJ_opinion_dynamics} as shown in Fig. \ref{Fig:expm_1_sim}. The agents $3,4$ and $5$ form an opinion cluster, agents $1$ and $6$ form another cluster and the agent $2$ forms the third cluster. Next, we modify $\mathcal{G}$ by adding edge $(1,5)$ with weight $0.5$ (and reducing the weight of $(4,5)$ to $0.5$ to ensure $W$ remains row-stochastic) as shown in Fig.\ref{Fig:modified_g}. For the same stubbornness and the initial opinions, Fig. \ref{Fig:final_op_modified_g} shows that node $5$ forms a new cluster while the rest belong to the same opinion clusters. 
   
\end{expm}

Example \ref{Example-1} demonstrates that the underlying topology has a significant impact on formation of opinion clusters in the network. Motivated by this, our objective is to establish the topology-based conditions that steer a subgroup of agents in the network to form an opinion cluster in the FJ framework. 
\section{The Relation of the final opinions with Network topology}
\label{sec:Kron_Red}

In this section, we present a framework to relate the final opinions of agents with the underlying network topology. Consider a digraph $\mathcal{G}$ consisting of $n$ agents with opinions evolving under the FJ model \eqref{eq:FJ_opinion_dynamics}. Let the network have $m$ stubborn agents. 
The final opinions satisfy the equations: $\mathbf{x}^*=(I- \beta)W \mathbf{x}^*+\beta \mathbf{x}(0)$. We can rewrite the above-mentioned linear equations as follows:
\begin{align}
        \underbrace{\begin{bmatrix}
         (I-(I-
    \beta)W ) &  -\eta &  \\
    \mathbb{0} &\mathbb{0}&
        \end{bmatrix}}_R \ \underbrace{\begin{bmatrix}
         \mathbf{x}^* \\ \mathbf{x}_s(0) 
        \end{bmatrix}}_z=\mathbb{0}
        \label{eqn:R_matrix}
    \end{align}
 
where $\mathbf{x}_s=[x_{s_1}(0),...,x_{s_m}(0)]$ and $s_1,s_2,...,s_m\in[m]$ are the labels of the stubborn agents. The matrix $\eta=[\eta_{ij}] \in \mathbb{R}^{n \times m}$ has entry $\eta_{{i}j}=\beta_{i}$ if the stubborn agent $i$ is labelled as $s_j$, otherwise $\eta_{ij}=0$ for $i\in[n],j\in[m]$. Henceforth, we represent eqn. \eqref{eqn:R_matrix} as $Rz=\mathbb{0}$ where $R=[r_{i,j}]\in \mathbb{R}^{(n+m) \times (n+m)}$. 

\begin{remark}
   The following are the salient properties of $R$: (a) each row sum in $R$ is $0$, (b) the diagonal entries are non-negative while the off-diagonal entries are non-positive. Thus, $R$ is a \textit{Laplacian matrix} under the Defn. 6.3 in \cite{bullo}. 
\end{remark}

 The digraph $\mathcal{G}(R)$ can be induced from $R$ if an edge $(i,j)$
exists in $\mathcal{G}(R)$ when $r_{ji}<0$.  
The associated graph $\mathcal{G}(R)$ has $n+m$ nodes. 
Being consistent with the indexing of nodes in $\mathcal{G}
$, the first $n$ nodes in $\mathcal{G}(R)$ are associated with $x_i^*$ for $i\in [n]$ and the nodes $n+1$ to $n+m$ are associated with the initial states of the $m$ stubborn agents. 
Note that each node $n+i$ forms a source in $\mathcal{G}(R)$ for $i\in[m]$. The source $n+i$ has a single outgoing edge to the node associated with the final opinion of the corresponding stubborn agent. For example, in Fig. \ref{fig:G(R)}, the initial opinion of stubborn agent $2$ is associated with node $n+1$ in $\mathcal{G}(R)$, thus, $n+1$ only has the outgoing edge $(n+1,2)$.

\begin{expm}
Consider the network $\mathcal{G}$ shown in Fig. \ref{Fig:expm_1_graph} with stubborn agents $2$ and $6$. The network $\mathcal{G}(R)$ for the matrix $R$ derived from eqn. \eqref{eqn:R_matrix} is shown in Fig. \ref{fig:G(R)}. The nodes in  $\mathcal{G}(R)$ that represent the initial opinion of the stubborn agents are highlighted in \textit{turquoise}.    
\begin{figure}[h]
    \centering
    \includegraphics[width=0.4\linewidth]{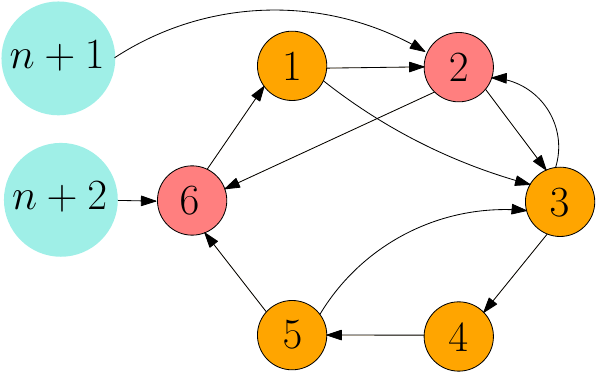}
    \caption{The network $\mathcal{G}(R)$ derived from $\mathcal{G}$.}
    \label{fig:G(R)}
\end{figure}
\end{expm}

\begin{remark}
\label{rem:suitable_KR}
Since $R$ is a Laplacian matrix, it can be reduced by Kron reduction for a suitable choice of the subset of nodes $\alpha$ (defined in Sec. \ref{subsec:KR}). Consider a network $\mathcal{G}$ without any oblivious agents. For such a network, each node in $[n]$ has a path in $\mathcal{G}(R)$ from at least one source $\{n+1,...,n+m\}$. Thus, by Statement 1) in Lemma \ref{lm:Basic_properties}, the Schur complement
$R/ \alpha^c$ is well-defined for any  $\alpha$ such that 
$\{n+1,...,n+m\} \subseteq \alpha \subseteq [n+m]$. If $\mathcal{G}$ has an oblivious agent, then the sources $\{n+1,...,n+m\}$ do not have a path to this agent. Thus, we define $\alpha$ to include the oblivious agents belonging to each iSCC along with sources $\{n+1,...,n+m\}$. For this choice of $\alpha$, each $i\in \alpha^c$ has a path in $\mathcal{G}(R)$ from a node $j\in \alpha$. Thus, Statement 1) in Lemma \ref{lm:Basic_properties} holds.

\end{remark}

Importantly, eliminating the states associated with nodes in $\alpha^c$ in eqn. \eqref{eqn:R_matrix}, reduces it to 
\begin{align}
\label{eqn:reduced_LE}
   R/ \alpha^c .\begin{bmatrix}
       \mathbf{x}^*[\omega] & \mathbf{x}_s(0)]
   \end{bmatrix}^T=\mathbb{0} 
\end{align}
  
where $\omega=\alpha \setminus \{n+1,...,n+m\}$.
Thus, the Kron reduction technique is useful in determining the relations between the final opinions of a subgroup of nodes in $\mathcal{G}$ and quantifying their dependence on the influential agents (stubborn agents and oblivious agents in iSCC). Using the Kron reduction technique, we present topology-based conditions under which the agents in a subgroup have equal final opinions and form an opinion cluster. 

\section{Topology-based opinion clustering}
\label{sec:OC}

Consider a weakly connected network $\mathcal{G}$ with opinions evolving under the FJ model \eqref{eq:FJ_opinion_dynamics}. Let $q$ be a non-influential agent in $\mathcal{G}$. Then, it is simple to see that the final opinion $x_q^*$ depends on the initial opinions of influential agents that have a path to $q$. Let $\mathcal{I}_q \subseteq \mathcal{I}$ be the set of all influential agents that have a directed path to $q$. Now, we define an LTP agent in terms of paths from agents in $\mathcal{I}_q$ to $q$.

\begin{definition}
\label{defn:LTP} An agent $p\in \mathcal{V}$ is called an \textbf{LTP agent} if the following holds:
\begin{enumerate}
    \item[(i)] there exists an agent $q\in \mathcal{V}\setminus \mathcal{I}$ such that every path in $\mathcal{G}$ from each influential agent $s\in \mathcal{I}_q$ to $q$ (where $p \neq q$) traverses $p$,
    \item[(ii)]  {additionally, if $p \in  \mathcal{V}\setminus \mathcal{I}$, an agent $c\in \mathcal{V}$ such that all paths from each agent $s \in \mathcal{I}_p$ to $p$ traverse $c$ must not exist.}
\end{enumerate}
For an LTP agent $p$, the non-influential agent $q$ satisfying condition (i) is said to be \textbf{\textit{persuaded by}} $p$.
\end{definition}

The set of all non-influential agents persuaded by an LTP agent $p$ is denoted as $\mathcal{N}_p$. Note that if $p\in \mathcal{I}$ is an influential agent, condition (i) alone ensures that $p$ is an LTP agent. The condition (ii) in Defn. \ref{defn:LTP} is required to be satisfied only when $p$ is non-influential. It implies that if a non-influential agent $p$  is persuaded by an LTP agent $c$ (\textit{i.e} $p\in \mathcal{N}_{c}$), then $p$ itself cannot be an LTP agent.

The following example illustrates the identification of an LTP agent and the agent(s) it persuades.

\begin{figure}[h]
    \centering
 \begin{subfigure}{0.23\textwidth}
        \centering
    \includegraphics[width=0.65\linewidth]{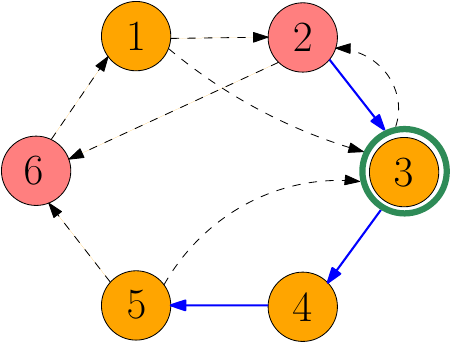}
    \caption{Paths from $2$ to $4$ (and $5$) traverse $3$}
    \label{Fig:LTP_expm_1}
    \end{subfigure}
    \hfill
    \begin{subfigure}{0.23\textwidth}
        \centering
    \includegraphics[width=0.65\linewidth]{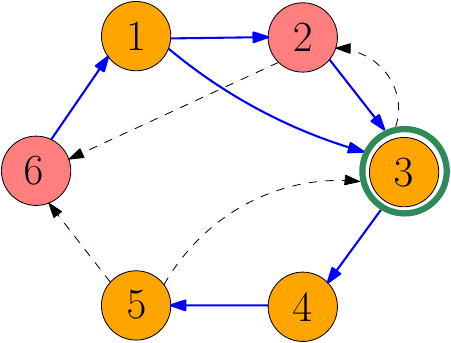}
    \caption{Paths from $6$ to $4$ (and $5$) traverse $3$}
       \label{Fig:LTP_expm_2}
    \end{subfigure}
    \caption{Node $3$ is an LTP agent and it persuades $4$ and $5$ as each path from the stubborn agents traverse $3$.}
    \label{fig:LTP_example}
\end{figure}

\begin{expm}
 Consider the network $\mathcal{G}$ shown in Fig. \ref{Fig:expm_1_graph}. Since $\mathcal{G}$ is strongly connected, only the stubborn agents $2$ and $6$ are influential and have a path to all agents. Fig \ref{Fig:LTP_expm_1} demonstrates that each path from $2$ to $4$ (and $5$) in $\mathcal{G}$ traverses through $3$. Moreover, the paths from $6$ to $4$ (and $5$) also traverse through $3$ as displayed in  Fig. \ref{Fig:LTP_expm_2}. Since the paths from $2$ and $6$ to $3$ do not traverse any common node. Thus, $3$ is an LTP agent and $\mathcal{N}_3=\{4,5\}$. Similarly, $6$ is also an LTP agent and $\mathcal{N}_6=\{1\}$. 

 {Note that each path from $2$ and $6$ to $5$ traverses $4$. Hence, $4$ satisfies condition 1) in Defn. \ref{defn:LTP}. However, since $4 \in \mathcal{N}_3$, it is not an LTP agent.}
\end{expm}

Suppose a network $\mathcal{G}$ has multiple LTP agents. The following question naturally arises: \textit{Can an agent be persuaded by two different LTP agents?} The following Lemma presents an answer to this question. 
\begin{lemma}
\label{lm:disjoint_partition}
Consider a network $\mathcal{G}$ and let $\mathcal{L}$ denote the set of LTP agents in $\mathcal{G}$. Then, $\mathcal{N}_i\cap \mathcal{N}_j=\emptyset$ for any distinct $i,j\in \mathcal{L}$.
\end{lemma}
\begin{proof}
Suppose the contrary is true and there exists a node $q\in \mathcal{N}_i \cap\mathcal{N}_j$. Then, by definition, each path in $\mathcal{G}$ from every influential agent in $\mathcal{I}_q$ to $q$ must traverse both $i$ and $j$. The following scenarios can occur:
\begin{enumerate}
    \item[C1] Every path from $s$ to $q$ traverses $i$ before $j$ for all $s\in \mathcal{I}_q$.
     \item[C2] Every path from $s$ to $q$ traverses $j$ before $i$ for all $s\in \mathcal{I}_q$. 
    \item[C3] Consider $s,u \in \mathcal{I}_q$. There exists a path from $s$ to $q$ that traverses $i$ before $j$, and there exists another path from $u$ to $q$ that traverses $j$ before $i$. Here, $s$ and $u$ can be the same node as well.
\end{enumerate}
Under C1, every path from an agent in $\mathcal{I}_q$ to $j$ also traverses $i$. Note that $j$ must be a non-influential agent under C1; otherwise $i$ will not be an LTP agent (as a path $j \to q$ exists that does not traverse $i$).
Since $j$ has a path to $q$, it follows that $\mathcal{I}_j \subseteq \mathcal{I}_q$, and hence every path from each node in $\mathcal{I}_j$ to $j$ traverses $i$.
Thus, $j \in \mathcal{N}_i$, so $j$ cannot be an LTP agent. Equivalently, under C2, we obtain $i \in \mathcal{N}_j$, so $i$ is not an LTP agent. Hence, both C1 and C2 result in contradiction.

Under C3, there exists a path $P_1$ from $s$ to $q$ that traverses $i$ before $j$ and another path $P_2$ from $u$ to $q$ that traverses $j$ before $i$. We can represent $P_1$ as $s \to i \to j \to q$ and $P_2$ as $u \to j \to i \to q$. By definition, a path does not contain repeated nodes. It follows from $P_1$ that a path $j \to q$ exists that does not traverse $i$. Further, we know from $P_2$ that a path from $u \to j$ does not traverse $i$. Therefore, a path from $u \to q$ exists that does not traverse $i$. Consequently, $q\notin \mathcal{N}_i$ which contradicts our assumption that $q\in \mathcal{N}_i \cap\mathcal{N}_j$. Similarly, we can show that a path from $s\to q$ exists that does not traverse $j$. 
 Since all of the conditions: C1, C2 and C3, lead to a contradiction, it means that the assumption $q \in \mathcal{N}_i \cap \mathcal{N}_j$ does not hold. \end{proof}

Lemma \ref{lm:disjoint_partition} establishes that an LTP agent and the agents it persuades form a disjoint group in the network. However, there can be certain agents who are neither an LTP agent nor are persuaded by any LTP agent. Excluding them, we can uniquely partition the remaining agents into disjoint groups, with each group consisting of one LTP agent and its persuaded agents. The following results presents the relation between the final opinions of agents within each such group. 
\begin{thm}
\label{thm:WC}
    Consider a network $\mathcal{G}$ of $n$ agents with $m$ stubborn agents. The opinions of agents in $\mathcal{G}$ are governed by the FJ model \eqref{eq:FJ_opinion_dynamics}. If each iSCC in $\mathcal{G}$ composed of the oblivious agents is aperiodic, then
    final opinions of an LTP agent $p$ and nodes in set $\mathcal{N}_p$ form an opinion cluster.
\end{thm} 

The proof of Theorem \ref{thm:WC} is given in the Appendix.

Theorem \ref{thm:WC} establishes that an LTP agent $p$ and the agents in $\mathcal{N}_p$  collectively form an opinion cluster in the final opinion. Note that the existence of an LTP agent in a network depends only on paths from the influential agents; hence, it is a purely graph topological property. Consequently, an LTP agent $p$ and $\mathcal{N}_p$ form an opinion cluster independent of the edge weights and the stubbornness of the stubborn agents. 

 Example \ref{Example-1} illustrates the result in Theorem \ref{thm:WC} as the LTP agent $3$ forms an opinion cluster with agents in $\mathcal{N}_3=\{4,5\}$ and LTP agent $6$ forms another opinion cluster with agent in $\mathcal{N}_6=\{1\}$. Additionally, it explains the formation of a new opinion cluster for the modified graph $\mathcal{\hat{G}}$. This occurs 
because the addition of edge $(1,5)$ results in a path
from $6$ to $5$ that does not traverse $3$. Thus, $5\notin\mathcal{N}_3$ and it does not form an opinion cluster with $3$ and $4$, as shown in Fig. \ref{Fig:final_op_modified_g}.

 By definition, if an LTP agent $p$ is oblivious, then the agents in $\mathcal{N}_p$ are also oblivious. Consequently, it follows from Theorem \ref{thm:WC} that a set of non-influential oblivious agents can also form an opinion cluster. Since the opinion of an oblivious agent evolves according to DeGroot model and is not affected by stubborn behaviour, an LTP agent ensures the formation of opinion clusters even in DeGroot framework. Note that since nodes in $\mathcal{N}_q$ are always non-influential, the desired opinion clusters can be formed in DeGroot framework only in weakly connected networks.

\begin{remark}
In \cite{yao2022cluster}, a network with at most one stubborn agent in each cycle is shown to form opinion clusters under the FJ model.
Theorem \ref{thm:WC} generalises the existing results by presenting LTP-based conditions to form an opinion cluster in any arbitrary digraph. 

Further, in \cite{yao2022cluster}, each opinion cluster is formed by a group that contains at least one influential agent: a stubborn agent with non-stubborn agents or the oblivious agents in an iSCC. In contrast, Theorem \ref{thm:WC} demonstrates that a group composed only of non-influential agents can also form an opinion cluster provided that one of them is an LTP agent and the rest are persuaded by it.
\end{remark}

\begin{figure}[h]
    \centering
 \begin{subfigure}{0.21\textwidth}
        \centering
    \includegraphics[width=1\linewidth]{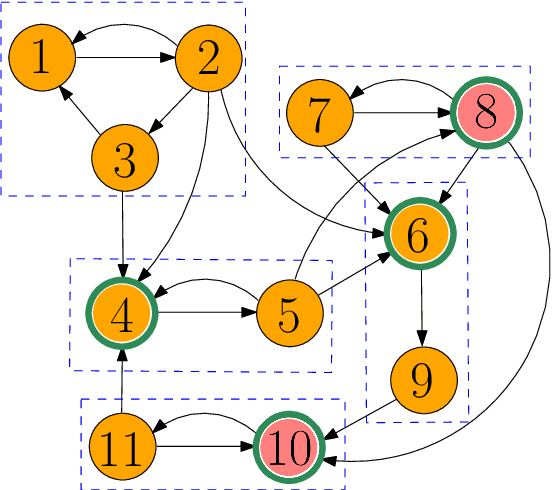}
    \caption{Network $\mathcal{G}$}
    \label{Fig:Network_with_oblivious_ag}
    \end{subfigure}
    \hfill
    \begin{subfigure}{0.25\textwidth}
        \centering
    \includegraphics[width=1\linewidth]{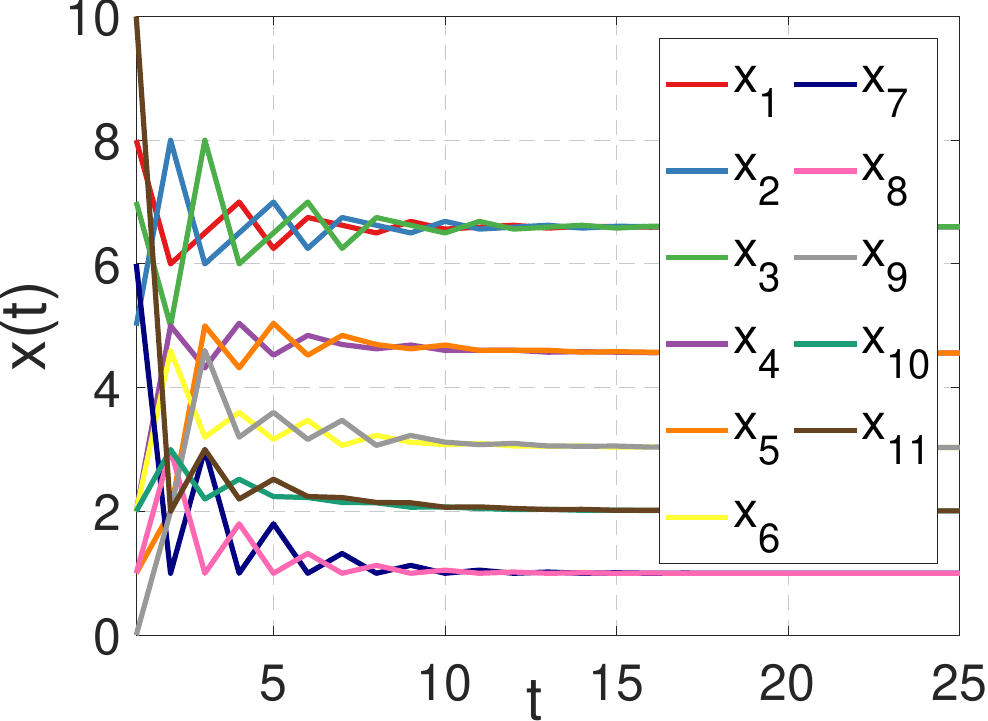}
    \caption{Formation of desired opinion clusters}
       \label{fig:desired_clustering}
    \end{subfigure}
    \caption{Suitable design of $\mathcal{G}$ for desired opinion clusters.}
    \label{Fig:final_op_and_stubbornness}
\end{figure}

Next, we illustrate the topological conditions presented in Theorem \ref{thm:WC} using the following example. 
\begin{expm}
\label{expm:desired_clusering}
Consider the network $\mathcal{G}$ with stubborn agents $8$ and $10$ in Fig. \ref{Fig:Network_with_oblivious_ag}. 
The agents $1,2$ and $3$ are oblivious agents which together form an iSCC (highlighted by a dashed red circle); the remaining agents are non-oblivious. The non-oblivious agents in $\mathcal{G}$ form $4$ disjoint groups highlighted by the green dashed boxes in Fig. \ref{Fig:Network_with_oblivious_ag}. Each 
group has an LTP agent (the node with green boundary) and the remaining agents in the group are persuaded by it. For initial conditions from a uniform distribution over $[0,10]$ and stubbornness values in $[0,1)$, the opinions of agents in $\mathcal{G}$ under the FJ model form $5$ opinion clusters as shown in Fig. \ref{fig:desired_clustering}. One cluster is formed by the oblivious agents in the iSCC and the rest are formed by the $4$ disjoint groups of agents; each composed of an LTP agent and those persuaded by it. 
\end{expm}

\section{Conclusions}
\label{Sec:conc}
The FJ model incorporates the individual biases as stubborn behaviour in the averaging-based opinion models. Due to varied biases, even the individuals in a closely connected group hold diverse opinions. Even then, a subgroup of individuals can still reach consensus, thereby, forming an opinion cluster. This work examines the role of the underlying network topology in the emergence of these clusters. To begin with, we define a topologically special agent called an LTP agent, and a set of non-influential agents that it persuades. Every path from an influential agent (a stubborn agent or an oblivious agent in an iSCC) to the persuaded agent always traverses the LTP agent. Using Kron reduction, we establish in Theorem \ref{thm:WC} that the final opinions of an LTP agent and the set of agents it persuades are equal. Thus, they form an opinion cluster in any arbitrarily-connected digraph under the FJ framework. Interestingly, such LTP agents lead to the formation of opinion clusters even under the DeGroot model (see Theorem \ref{thm:WC}).

Some additional key insights from our results are: (i) The presence of LTP agents always results in opinion clusters, independent of the edge-weights in the network and the stubbornness of the agents. (ii) Notably, an LTP agent may or may not be influential, but still shapes the final opinions of the agents it persuades. (iii) The notion of LTP agents generalises the topology-based conditions for opinion clustering in \cite{yao2022cluster} to any arbitrary digraphs. (iv) By suitably placing the LTP agents and designing the network topology, we achieve any desired opinion clustering as demonstrated in Example \ref{expm:desired_clusering}. 

In future, we plan to examine the potential of LTP agents in desirably shaping the final opinions in social networks.

\section{Appendix}
Before moving to the proof of Theorem \ref{thm:WC}, the following Lemma needs to be stated.

\begin{lemma}
\label{lemma:lm-inverse}
Consider a weakly connected network $\mathcal{G}$ with $m\geq 1$ stubborn agents and matrix $R$ derived using eqn. \eqref{eqn:R_matrix}. 
Let $\alpha$ be a set containing $\{n+1,....,n+m\}$ and all the oblivious agents in the iSCCs in $\mathcal{G}$ and $\alpha^c=[n+m]\setminus \alpha$ such that $\alpha^c\neq \phi$. Then,

\begin{align}
\label{eqn:identity-1}
   (R[\alpha^c])^{-1}=\sum_{k=0}^{\infty} \big((I-\beta[\alpha^c])W[\alpha^c]\big)^{k} 
\end{align}

\end{lemma}
\begin{proof}
By definition of $\alpha$, we get $\alpha^c \subseteq [n]$. From the construction of $R$, it follows that $R[\alpha^c]=I-(I-\beta[\alpha^c])W[\alpha^c]$. Let $M=(I-\beta[\alpha^c])W[\alpha^c]$. Since $W$ is row-stochastic, $M$ satisfies $M\mathbb{1}_n=(I-\beta)\mathbb{1}_n\leq \mathbb{1}_n$. Hence, the row-sum of any row in $M$ can be at most $1$. By the Gersgorin disk theorem, the spectrum of $M$ is a subset of a unit disk in the complex plane (Theorem 2.9 in \cite{bullo}). 

Further, since $M$ is a non-negative matrix, by the Perron-Frobenius Theorem, there exists a real eigenvalue $\lambda_M$  such that $\lambda_M\geq |\mu|\geq 0$, where $\mu$ are the remaining eigenvalues of $M$ (Theorem 2.12 in \cite{bullo}). Thus, $\lambda_M\in[0,1]$. 

Further, as discussed in Remark \ref{rem:suitable_KR}, the Schur complement is well defined for any $\alpha$ containing $\{n+1,....,n+m\}$ and the oblivious agents in iSCCs of $\mathcal{G}$. Thus, $R[\alpha^c]$ is invertible. This means that none of the eigenvalues of $M$ is equal to $1$. Thus, $\lambda_M<1$ and eqn. \eqref{eqn:identity-1} follows by Neumann series.
\end{proof}

\textbf{Proof of Theorem 1:}
Let $q\in \mathcal{N}_p$. To prove that the final opinion $x_p^*=x_q^*$, we reduce the matrix $R$ in eqn. \eqref{eqn:R_matrix} using the  {Kron Reduction} and examine the relation between $x_p^*$ and $x_q^*$ in the reduced set of linear equations \eqref{eqn:reduced_LE}. 
 {Let the first $u$  agents in $\mathcal{G}$ be oblivious agents that form the iSCCs. The oblivious agents in each iSCC in $\mathcal{G}$ have consensus \cite{TIAN2018213}. Hence, we consider $q$ is not an oblivious agent in an iSCC; otherwise, $p$ and $q$ belong to an iSCC that is known to achieve consensus.}

Consider $\alpha$ to be $\{p,q,1,...,u,n+1,...,n+m\}$.
From Remark \ref{rem:suitable_KR}, we know that the Kron Reduction $R/ \alpha^c$ is well-defined.
Next, we evaluate $R/ \alpha^c$ given by eqn. \eqref{eqn:Schur_complement}. 
The following structural property of the associated network $\mathcal{G}(R)$ will be used in determining $R/ \alpha^c$:

\begin{enumerate}
     \item[\textbf{P1}:] Each walk that exists from $s\in \mathcal{I}_q$ to $q$ in $\mathcal{G}$ traverses $p$ for all $s\in \mathcal{I}_q$. Thus, by construction of $\mathcal{G}(R)$, each walk that exists from $j$ to $q$ in $\mathcal{G}(R)$ will traverse $p$ for all $j\in \{[u],n+1,...,n+m\}$.
\end{enumerate}
Property \textbf{P1} implies that $r_{q,j}=0$ for all  {$j\in\{[u],n+1,...,n+m\}$} because node $j$ cannot be an in-neighbour of $q$ in $\mathcal{G}(R)$. Additionally, the nodes $n+1,...,n+m$ in $\mathcal{G}(R)$ are sources. Thus, the corresponding rows of $R$ have all entries as zero. Therefore, $R[\alpha]$ takes the following form:
$
{\begin{bmatrix}
r_{p,p} & r_{p,q} & R[p,[u]] & r_{p,n+1} & ... & r_{p,n+m}  \\
r_{q,p} & r_{q,q} & \mathbb{0} &  0 & ... & 0 &   \\
 \mathbb{0}& \mathbb{0}& R[[u]] &\mathbb{0} &\cdots & \mathbb{0}\\
\mathbb{0}_{m \times 1} & \mathbb{0}_{m\times 1} & \mathbb{0}_{m \times 1} & \mathbb{0}_{m\times 1} & ... & \mathbb{0}_{m \times 1}
\end{bmatrix}}$. 

Next, we consider $Y=R[\alpha,\alpha^c](R[\alpha^c])^{-1}R[\alpha^c,\alpha]$.
From Lemma \ref{lemma:lm-inverse}, it follows that 
\begin{align}
\label{eqn:Y_structure}
    Y=R[\alpha,\alpha^c]\sum_{k=0}^{\infty} \big((I-\beta[\alpha^c])W[\alpha^c]\big)^{k}R[\alpha^c,\alpha]
\end{align}

Let $F:\alpha\to [ {u}+m+2]$
be a mapping such that $F(p)=1,F(q)=2$,  {$F(j)=j+2$ for $j\in [u]$,
 and $F(n+j)=u+2+j$} for $j\in[m]$. This 
mapping relates the agents in set $\alpha$ with the row (and column) indices of $Y$ (and $R/\alpha^c$). Then, from eqn. \eqref{eqn:Y_structure}, the entry $y_{F(i)F(j)}$ of $Y$ is non-zero only if there exists a walk in $\mathcal{G}(R)$ from $j$ to $i$ such that each node on this walk (except $i$ and $j$) belongs to $\alpha^c$. By property \textbf{P1}), $y_{2F(j)}=0$ for all  $j \in  [u] \cup \{n+1,...,n+m\}$ because each walk from corresponding $j$ to $q$ in $\mathcal{G}(R)$ always passes through $p \in \alpha$. Consequently, $R/\alpha^c=R[\alpha]-Y$ has the form:

$ \begin{bmatrix}
r_{1,1}^1 & r_{1,2}^1 & \Delta_1 & r_{1,u+3}^1 & ... & r_{1,u+m+2}^1 \\
r_{2,1}^1 & r_{2,2}^1 & \mathbb{0}_{u \times 1} & 0 & ... & 0 \\
 \mathbb{0}_{u \times 1}& \mathbb{0}_{u \times 1}& R[[u]] & \mathbb{0}_{u \times 1} &\cdots & \mathbb{0}_{u \times 1}\\
\mathbb{0}_{m\times 1} & \mathbb{0}_{m\times 1} &  \mathbb{0}_{m\times 1}& \mathbb{0}_{m \times 1} &... & \mathbb{0}_{m \times 1}
\end{bmatrix}$

where $r_{i,j}^1$ gives the entry of matrix $R / \alpha^c$ for $i,j\in [m+u+2]$.  {The vector $\Delta_1=[r_{1,3}^1,r_{1,4}^1,...,r_{1,u+2}^1]$ highlights the corresponding non-zero entries.}
From Lemma \ref{lm:Basic_properties}, we know that $R/ \alpha^c$ is also a Laplacian matrix and the row-sum of each of its rows equals $0$. Since the second row of $R/\alpha^c$ has only two non-zero entries, thus, $r_{2,1}^1=-r_{2,2}^1$. Moreover, by eliminating the states $\alpha^c$ in the steady state eqns. \eqref{eqn:R_matrix}, we get $x_p^*=x_q^*$. Since $q$ is any node in $\mathcal{N}_p$, this holds for all nodes in $\mathcal{N}_p$.
\hfill $\blacksquare$.

\bibliographystyle{IEEEtran}
\bibliography{IEEEabrv,references}
\end{document}